\documentclass[prb,superscriptaddress,twocolumn,floats,floatfix,]{revtex4}

\usepackage{graphicx}
\usepackage{gensymb}
\usepackage{amsmath}
\begin{document}

\title{Orbital and valley state spectra of a few-electron silicon quantum dot}

\author{C. H. Yang}
\email{henry.yang@unsw.edu.au}
\author{W. H. Lim}
\author{N. S. Lai}
\email[Present address: University College of Technology and Innovation, Malaysia]{}
\author{A. Rossi}
\author{A. Morello}
\author{A. S. Dzurak}
\affiliation{ARC Centre of Excellence for Quantum Computation and Communication Technology, School of Electrical Engineering \& Telecommunications, The University of New South Wales, Sydney 2052, Australia}

\date{\today}

\begin{abstract}
Understanding interactions between orbital and valley quantum states in silicon nanodevices is crucial in assessing the prospects of spin-based qubits. We study the energy spectra of a few-electron silicon metal-oxide-semiconductor quantum dot using dynamic charge sensing and pulsed-voltage spectroscopy. The occupancy of the quantum dot is probed down to the single electron level using a nearby single-electron transistor as a charge sensor. The energy of the first orbital excited state is found to decrease rapidly as the electron occupancy increases from $N$ = 1 to 4. By monitoring the sequential spin filling of the dot we extract a valley splitting of $\sim$230$~\mu$eV, irrespective of electron number. This indicates that favourable conditions for qubit operation are in place in the few-electron regime.
\end{abstract}

\maketitle

\section{Introduction}
\label{intro}
Silicon-based quantum dots (QD) are recognized as a promising system for the implementation of solid state quantum computing~\cite{prance,Simmons2011,Maune,lai,Angus2007}. While such devices are appealing because they can be engineered to show long coherence times~\cite{Witzel,tyry1,tyry2}, there also exist complications for the manipulation of quantum states which arise from the multi-valley conduction band present in bulk silicon~\cite{Ando1982}. In fact, valley degeneracy is typically lifted in nanodevices due to interfacial stress/strain and electrostatic confinement~\cite{Lim2011,Borselli2011}. Hence, understanding valley physics and its perturbing effects on spin relaxation process has become paramount for the practical realization of spin-based quantum bits (qubits). In order to assess whether a quantum device shows valley splitting compatible with use in spin-based quantum computing, orbital and valley spectra must be extracted. In particular, it is of interest to probe the relative magnitude of the valley and orbital energy spacing to determine the degree of mixing of these states~\cite{friesen} and ultimately identify the most appropriate strategy for qubit operation~\cite{Culcer2009}.\\\indent
In this work, we use pulsed-voltage spectroscopy to investigate the excitation energy spectra of a nearly-closed silicon metal-oxide-semiconductor (MOS) QD in the few-electron regime ($N\leq 4$). In a magnetic field we observe Zeeman shifts of the ground state, valley excited state, and first orbital excited state that allow us to determine the sequential spin filling of the $N$-electron states. Interestingly, the energy of the first orbital excited state is seen to decrease rapidly as the dot occupancy increases, whereas a valley splitting of $\sim$230~$\mu$eV is found not to depend on electron number. This has the significance of guaranteeing the existence of a well-defined spin-1/2 qubit Hilbert space, sufficiently separated from higher-energy excitations.
\begin{figure}[b]
\includegraphics[width=7.5cm]{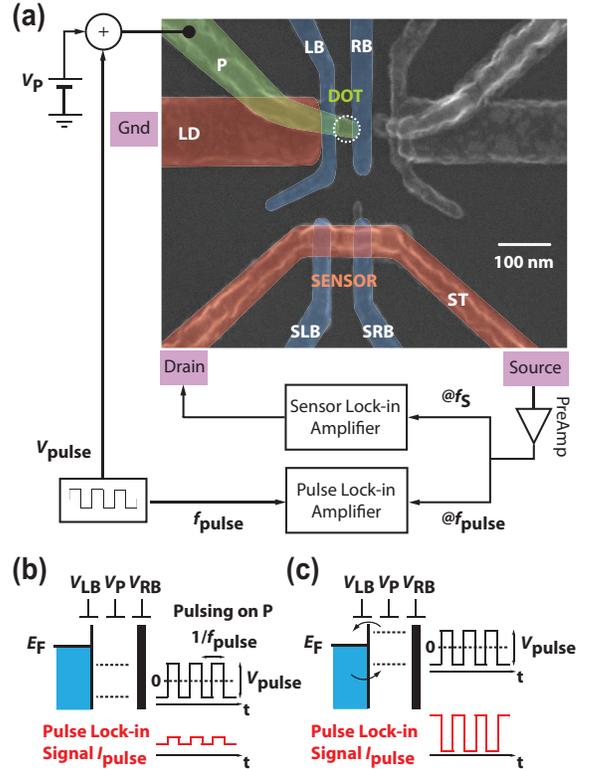}
\caption{(Color online)(a) SEM image of a quantum dot integrated with a SET and the measurement set-up. Only colored gates are used in the experiments. (b)(c) Schematic energy diagrams during pulsing of gate P with two different $V_\textrm{P}$ offsets.}
\label{sem}
\end{figure}
\section{Experimental methods}
\label{methods}
Fig.~\ref{sem}(a) shows a scanning electron microscope (SEM) image of a device identical to the one under study; in the experiments presented here, only the (false) coloured gates are used. The device works as a single-lead quantum dot (upper structure) integrated with a single-electron transistor sensor (lower structure). It was fabricated using multi-layer Al-Al$_2$O$_3$-Al gate stack technology and the detailed fabrication process has been described in Refs.~\cite{Angus2007,Lim2009b}. A quantum dot is formed between the left and right barrier gates (LB, RB) and is independently controlled by the plunger gate (P). By applying a positive voltage on the lead gate (LD), an electron accumulation reservoir is induced at the Si/SiO$_2$ interface. Electrons from the reservoir can then tunnel on and off the quantum dot through the left barrier. Next to this single-lead quantum dot is a single-electron transistor (SET) sensor (S) which is used to detect the electron tunneling events to/from the quantum dot. The SET top gate (ST) forms an electron channel from source to drain and the two underlying barrier gates (SLB, SRB) create tunnel barriers, forming an electron island in between. The measurements were performed in a dilution refrigerator with base temperature of $\sim$50~mK.\\\indent
In order to characterize the excitation spectrum~\cite{Klimeck1994} of the QD, we make use of the technique developed by Elzerman and co-workers~\cite{Elzerman2004} which combines charge detection and gate pulsing. We apply an $ac$ excitation voltage of 200~$\mu$V at $f_\textrm{S}$=173~Hz to the drain of the SET and monitor the sensor current $I_\textrm{S}$ locked-in to $f_\textrm{S}$ at the source through a low-noise room temperature preamplifier. Simultaneously, a train of voltage pulses with amplitude $V_\textrm{pulse}$ is applied to gate P, in addition to its dc voltage $V_\textrm{P}$, shifting the energy levels of the dot up and down. This pulse train modulates the sensor current via a cross-capacitance at a frequency $f_\textrm{pulse}$ and the resulting current $I_\textrm{pulse}$ at this frequency is measured with a second ($Pulse$) lock-in amplifier, as shown in Fig.~\ref{sem}(a). In order to maximize charge sensitivity in the detector, we employ dynamic compensation, as described in Yang \emph{et al}.~\cite{Yang2011}. We first tune the SET so that the sensor lock-in current $I_\textrm{S}$ is at the edge of a Coulomb peak, where the transconductance $dI_\textrm{S}$/$dV_\textrm{ST}$ is high (2~nS). We then monitor the charge state of the quantum dot using the sensor signal $I_\textrm{S}$ while the SET gate voltage $V_\textrm{ST}$ is dynamically adjusted in order to maintain an approximately constant sensor signal $I_\textrm{S}$. This makes our read-out signal virtually unaffected by slow charge drifts and random charge rearrangements, which would otherwise prevent optimal bias.

Fig.~\ref{sem}(b)(c) show schematic energy diagrams of electrons loading and unloading the quantum dot through the left barrier while the right barrier is raised high to completely cut off the channel. The electron tunnel rate is independently controlled by the left barrier gate voltage $V_\textrm{LB}$. In (b), both the high and low phase of the pulse are below the Fermi level $E_\textrm{F}$ of the electron reservoir. Hence, there is no change in electron occupancy of the dot and the pulse lock-in detection signal $I_\textrm{pulse}$ is small, due only to capacitive coupling between the plunger gate and the sensor. Conversely, in (c), an electron can tunnel into the quantum dot from the reservoir during the high phase of the pulse (low in potential) and tunnel off when the pulse phase is low (high in potential). This change of electron occupancy in the dot induces a much larger lock-in detection signal $I_\textrm{pulse}$.

 \begin{figure}[t]
  \includegraphics[width=7.5cm]{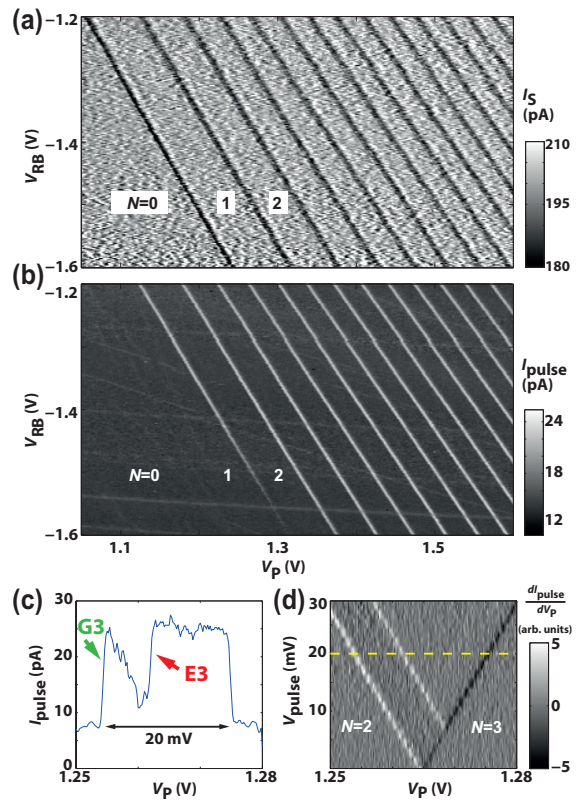}
 \caption{(Color online)
 (a) SET sensor lock-in current $I_\textrm{S}$ vs. $V_\textrm{P}$ and $V_\textrm{RB}$, mapping out the charge stability plot of the quantum dot. Dynamic compensation is applied to the SET sensor. (b) Measured pulse lock-in signal $I_\textrm{pulse}$ from SET sensor with a symmetric square wave of peak-to-peak voltage $V_\textrm{pulse}=2$~mV applied to dot plunger P at $f_\textrm{pulse}$=487~Hz. The first charge transition is not visible because the tunnel rate into/out of dot is too low with respect to $f_\textrm{pulse}$. (c) Lock-in detection signal, with $V_\textrm{pulse} = 20$~mV, at $f_\textrm{pulse}$=444~Hz for $N = 2\leftrightarrow3$ transition, extracted along the yellow dashed line in (d) at $V_\textrm{pulse}$ = 20~mV. The red arrow indicates the orbital excited state. (d) Derivative of the detection signal with respect to $V_\textrm{P}$. Orbital ground and excited states are observed at the loading edge as white parallel lines.}
 \label{chplot}
 \end{figure}
\section{QD occupancy}
\label{occupancy}
Fig.~\ref{chplot}(a) shows the measured QD charge stability map as a function of right barrier gate voltage $V_\textrm{RB}$ and plunger gate voltage $V_\textrm{P}$. As we reduce $V_\textrm{P}$, the number of electrons in the quantum dot is reduced one by one until we observe no more charge transitions in the stability map. The charging energy of the dot is around 10~meV for high occupancy and increases significantly (up to 20 meV) in the few-electron regime. This indicates that the dot size is dramatically affected by the number of electrons at low $N$ and strongly suggests that we have achieved the few-electron regime. In light of this, we believe that the last transition observed is likely to be ascribed to the last electron ($N=0$).  Simultaneously, we measure the pulse lock-in detection signal $I_\textrm{pulse}$ when a symmetric square wave with $V_\textrm{pulse}$ of 2~mV and $f_\textrm{pulse}$=487~Hz is applied to gate P, and the resulting stability map is plotted in Fig.~\ref{chplot}(b). Identical charge transitions are detected using $I_\textrm{pulse}$, but with an improved signal-to-noise ratio compared with the sensor signal $I_\textrm{S}$ plotted in Fig.~\ref{chplot}(a). This allows us to detect a number of extra transitions which are believed to originate from charge impurities nearby the sensor. Indeed, these features have different slopes with respect to the main ones and their signal strength is significantly reduced. This indicates that the originating charge displacement is elsewhere located in the substrate with respect to the gate-defined QD. 

Towards the one-electron limit, as the dot potential well becomes shallower and the tunnel barrier widens, the tunnel rate decreases. Here, the tunnel rate for the first electron falls significantly below 974~Hz (2$\times{}f_\textrm{pulse}$~Hz) and so the $N$=0$\leftrightarrow$1 transition is not visible in Fig.~\ref{chplot}(b). Although electron tunneling is allowed energetically (see Fig.~\ref{chplot}(a)), the high pulsing frequency used does not provide an electron with sufficient time to tunnel onto/off the dot. In this case, the tunneling time is longer than 1~ms.
\begin{figure}[]
\includegraphics[width=8.5cm]{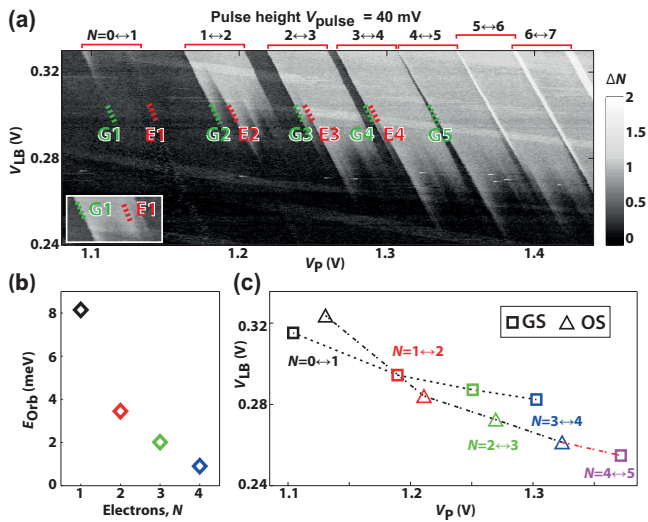}
\caption{(Color online)(a) Lock-in signal from sensor S with $V_\textrm{pulse}$=40~mV at $f_\textrm{pulse}$=487~Hz applied to gate P, as a function of $V_\textrm{P}$ and $V_\textrm{LB}$. Here we apply a linear scaling of the lock-in output to directly indicate the excess electron occupancy $\Delta{}N$ on the dot. INSET: different region focusing on the first transition shown to enhance visibility of both excited state and unloading edge. (b) Orbital excited-state energy of the first 4 electrons at $V_\textrm{LB}$=0.3~V.(c) Positions of $\Delta{}N$ = 0.5 for ground states (GS) and first excited states (OS). These points are extracted from panel (a) where the tunnel rate from the reservoir to the dot is around 974~Hz.}
\label{excite}
\end{figure}

In order to observe the excited states of the dot, we increase $V_\textrm{pulse}$ to 20~mV as shown in Fig.~\ref{chplot}(c-d) for the $N=2\leftrightarrow3$ transition. The pulse lock-in signal $I_\textrm{pulse}$ increases as soon as the three-electron ground state is pulsed below $E_\textrm{F}$ (green arrow in Fig.~\ref{chplot}(c)). The signal then decays slowly with increasing $V_\textrm{P}$ indicating a decrease in tunnel rate as the ground state moves away from resonance with the Fermi level. When the three-electron excited state is pulsed below $E_\textrm{F}$ the tunnel rate rises rapidly again, resulting in a second increase in the detection signal (red arrow in Fig.~\ref{chplot}(c)). When we increase $V_\textrm{P}$ further, so $N$ = 3 at both pulse levels, the signal falls low again since the tunneling events are energetically forbidden due to Coulomb blockade. Fig.~\ref{chplot}(d) shows the derivative $dI_\textrm{pulse}$/$dV_\textrm{P}$ of the pulse lock-in detection signal as a function of $V_\textrm{pulse}$ and $V_\textrm{P}$. The left white line represents the ground-state loading edge. The second white line in parallel with the ground state is the excited state loading edge, while the unloading edge appears as a black line. Here by loading (unloading) edge we refer to the potential configuration for which an electron can start to occupy an energy level in the upper phase of the pulse (not) being allowed to empty it during the lower phase.

\section{Orbital spectrum}
\label{orbit}
We next study in detail the orbital excited states for the first few electrons by increasing $V_\textrm{pulse}$ to 40~mV and plotting the charge stability diagram of the dot as a function of $V_\textrm{LB}$ and $V_\textrm{P}$ in Fig.~\ref{excite}(a). We identify these excited states as orbital states, as opposed to valley states, because their energies are much larger than the valley splitting in our device ($\sim$230~$\mu$eV), as we shall show in the next section. In Fig.~\ref{excite}(a), we have converted the pulse lock-in signal $I_\textrm{pulse}$ to a corresponding average change of electron occupancy $\Delta N$ in the dot using a simple linear map. We identify $\Delta N = 1$ as the level where $I_\textrm{pulse}$ saturates, which occurs when the tunnel rate significantly exceeds the pulse frequency $f_\textrm{pulse}$, as seen in Fig.~\ref{chplot}(c) at $V_\textrm{P}$ beyond the E3 riser. Note that $\Delta N = 2$ where two transitions overlap, as the pulsing level exceeds the charging energy allowing occupancy to change by two electrons. The ground states and the orbital excited states of the first four electron transitions are highlighted by green and red lines, respectively. We also observe a decrease in tunnel rate as we decrease $V_\textrm{LB}$ (thus increasing the barrier height), resulting in $\Delta N \rightarrow 0$ as the tunnel rate falls below $f_\textrm{pulse}$. In Fig.~\ref{excite}(b), we plot the first orbital excited state energy $E_\textrm{Orb}$ with respect to the electron number $N$, extracted from Fig.~\ref{excite}(a). Note that the energy conversion factor, $\alpha\sim$ = 0.3~eV/V, is derived from considerations relevant to valley splitting measurements, as discussed next. As $V_\textrm{P}$ is reduced, the dot becomes smaller leading to an increase in the orbital level spacing from 1~meV for $N = 4$, to 8~meV for $N = 1$. 

In Fig.~\ref{excite}(c), we plot the positions in gate voltage space ($V_\textrm{LB}$; $V_\textrm{P}$ )corresponding to $\Delta N = 0.5$ for the ground and orbital excited states $N$ = 1 to 5. These are the points where the tunnel rates are comparable to the pulse frequency $f_\textrm{pulse}$. As the dot occupancy increases, a greater barrier height (or lower $V_\textrm{LB}$) is required to maintain a constant tunnel rate. We note that both the ground states (GS) and the orbital excited states (OS) follow regular trend lines, except for the $N = 5$ ground state, which appears more like an orbital excited state. This is due to the ground orbital levels being full with electrons and starting to fill the next orbital.

\section{spin-valley interaction}
\label{valley}
\begin{figure}[]
\includegraphics[width=8.5cm]{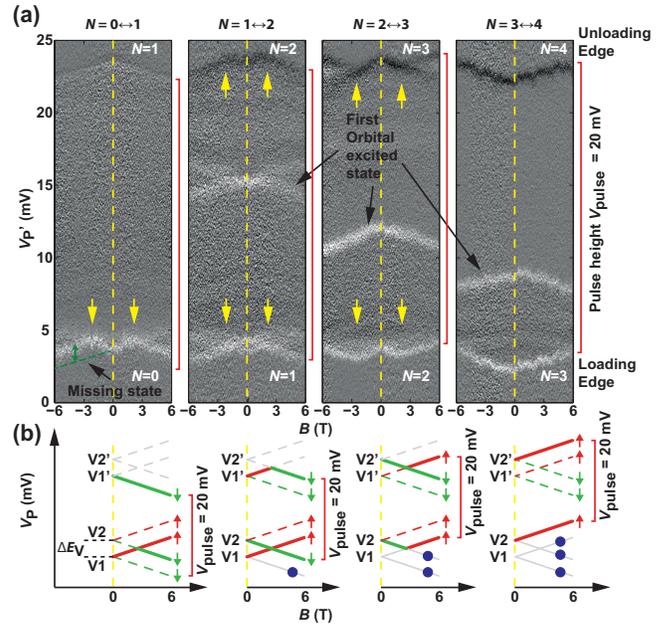}
\caption{(Color online)(a) Grey-scale plot of the QD differential occupancy ($\Delta^2N$) in magnetic field for the first 4 electron transitions at $V_\textrm{pulse}=$20~mV and $f_\textrm{pulse}$=487~Hz. Orbital excited states, Zeeman splittings and valley-orbit splitting are observed. $V_\textrm{P}'$ represents arbitrarily shifted $V_\textrm{P}$. (b) Schematic of the evolution of two non-degenerate valley eigenstates for positive $B$-field. V$_1$ (V$_1$') and V$_2$ (V$_2$') represent loading (unloading) edge states. Each valley state Zeeman splits into two levels at finite $B$-fields. The blue dots represent already-occupied electron states.}
\label{magnet}
\end{figure}

We now study the spin filling of the first four electrons into the valley states of the quantum dot by examining the Zeeman shifts of the ground and excited states in a magnetic field parallel to the oxide interface in the range -6~T $<B<$ 6~T, as shown in Fig.~\ref{magnet}(a). Here we set $V_\textrm{pulse}$ = 20~mV and plot the derivative of the pulse lock-in current, $dI_\textrm{pulse}$/$dV_\textrm{P}$. The data reveals a number of levels, including spin ground and excited states, for what appear to be two distinct valley levels, as we explain below.

We first consider the unloading edge of the charge transitions, which appear as dark lines in Fig.~\ref{magnet}(a). For the first transition ($N = 0\leftrightarrow1$), the unloading edge moves towards less positive $V_\textrm{P}$ with increasing magnetic field $|B|$, indicating that the first electron is spin-down $|$$\downarrow\rangle$. Assuming that the g-factor = 2 for electrons in silicon and using the Zeeman energy equation $\frac{1}{2}g\mu_B B$, where $\mu_B$ is the Bohr magneton, we fit the energy change of the unloading edge. This results in a plunger gate voltage to energy conversion factor, $\alpha\sim$ 0.3~eV/V. This is consistent with values for a similar quantum dot reported in Ref.~\cite{Lim2009b}. Note that the evaluation of the $\alpha$-factor from the Zeeman splitting is a robust approach in Si-based systems. Because of the weak spin-orbit coupling~\cite{roth60pr}, only very small  deviations (less than 1\%) of the g-factor from the free-electron value are observed in Si, as opposed to the case of III-V semiconductors where much larger deviations are found. Next, we analyze the unloading edge of the second transition ($N = 1\leftrightarrow2$). The energy level first increases with magnetic field for $|B|$ $<$ 2 T, indicating that a spin-up electron was unloaded, and then decreases with $|B|$ above $\sim$ 2 T. This indicates that the $N = 2$ ground state changes from a singlet $\frac{1}{\sqrt2}(\left|\downarrow\uparrow\right\rangle-\left|\uparrow\downarrow\right\rangle)$ at low $B$, to a triplet $\left|\downarrow\downarrow\right\rangle$ for $|B|$ $>$ 2 T. A similar kink at $|B|$ $\sim$ 2 T (marked by a yellow arrow) is observed for the third transition ($N = 2\leftrightarrow3$), although here the unloading edge moves downwards at low $|B|$ and then upwards at high $|B|$. Finally, the unloading edge of the fourth transition ($N = 3\leftrightarrow4$) rises linearly with $|B|$, consistent with a spin-up electron unloading for all magnetic fields up to 6 T.\\\indent
In order to explain these data, we use a model based on two non-degenerate valley states (V$_1$, V$_2$) whose spin-states depend on B field according to the Zeeman energy equation (Fig.~\ref{magnet}(b)). This model relies on the fact that bulk silicon has six degenerate valleys, but when confined to two dimensions, such as a Si/SiO$_2$ interface, these separate into four valleys with high effective mass and two lower-energy $\Gamma$-valleys~\cite{Ando1982}. The $\Gamma$-valleys are also non-degenerate with a splitting $\Delta E_\textrm{V} = E_\textrm{V2} - E_\textrm{V1}$ that is typically well below 1~meV. In previous ground-state magneto-spectroscopy studies on a similar dot we found  $\Delta E_\textrm{V}$ $\sim$ 0.1 meV~\cite{Lim2011}, which we note is much smaller than the first orbital excited state energies ($E_\textrm{Orb}$ = 1$-$8~meV) as long as dot occupancy is kept low ($N\leq 4$). From Fig.~\ref{magnet}(b), we see that the first electron fills the lower valley V$_1$ as spin-down for all $B$. When the second electron loads ($N = 1\leftrightarrow2$ transition) it will fill V$_1$ as spin-up at low values of $|B|$, forming a spin singlet $\frac{1}{\sqrt2}(\left|\downarrow\uparrow\right\rangle-\left|\uparrow\downarrow\right\rangle)$, however at large $|B|$, when the Zeeman energy exceeds the valley splitting $\Delta E_\textrm{V}$, it will preferentially load into the upper valley V$_2$ as spin-down, to form a triplet state $|$$\downarrow\downarrow\rangle$. Moving to the third electron, we see that it will fill as spin-down in valley V$_2$ for low $|B|$, and then as spin-up in valley V$_1$ once the Zeeman energy exceeds $\Delta E_\textrm{V}$. The fourth electron always fills as spin-up in valley V$_2$, unless there is a nearby orbital state (within the Zeeman energy), which is likely to be the case for higher dot occupancy than currently reported ($N> 4$). In Fig.~\ref{magnet}(a) we can clearly see the orbital excited states (white lines) between the loading and unloading edges. They predominantly move downwards with the Zeeman energy, implying that they load spin-down electrons.

We find that the level crossings in Fig.~\ref{magnet} (marked by yellow arrows) occur at $|B| \sim$ 2~T for $N = 1 .. 3$, indicating that the valley splitting $\Delta E_\textrm{V}= g\mu_B B \sim$ 230~$\mu$eV remains approximately constant for small electron occupancy. This value is roughly double the valley splitting we observed in a previous device~\cite{Lim2011}, most likely due to the larger electric fields employed in the present study~\cite{Saraiva2010}. The observation of a valley splitting that is independent of electron number confirms the prediction~\cite{Culcer2010} that the valley exchange Coulomb integral is negligible.

The ground-state loading edges (lower white lines in Fig.~\ref{magnet}(a)) for transitions 2 to 4 show the same trends in magnetic field as the unloading edges. However, for the first transition ($N=0\leftrightarrow1$) the ground state (marked by a green dashed line) is not visible. We know this state is present because the distance between the loading and unloading edge must equal $V_\textrm{pulse}$ = 20 mV. We believe this is due to a very low (loading) tunnel rate, which we previously observed for low electron occupancy in Fig.~\ref{excite}(a). Note that the observation of the spin excited state is compatible with the pulse frequency in use which is chosen to be higher than the relaxation rate. Indeed, the relaxation rate in the window of magnetic field applied is expected to be in the range 0.1~-~10 Hz~\cite{Simmons2011,Xiao2010}, whereas we use pulse sequences of few hundreds of Hz.  This accounts for the kink at 2~T formed by the lower valley spin up state and the upper valley spin down state. Finally, we consider the $N = 5$ ground state, which has a tunnel rate that follows the trend for the orbital excited states for $N$ = 1 to 4, as shown in Fig.~\ref{excite}(c). We can now understand this in terms of a shell structure, where the first four electrons fill the spin and valley states of the lowest orbital, and the fifth electron occupies the next available orbital level. This is consistent with a shell structure of $N = 4$ observed in other silicon quantum dots~\cite{Lim2011,Borselli2011}.

\section{conclusions}
\label{conclusions}
In conclusion, we have presented excited state spectroscopy of a nearly-closed silicon quantum dot using charge sensing and a pulsed-gating technique, thus enabling clear identification of the spin, valley and orbital states for the first four electrons. As the occupancy increased from $N$ = 1 to 4 electrons, we found that the valley splitting for the lowest orbital level remained approximately constant at 230$~\mu$eV, while the next orbital level energy decreased from 8~meV to 1~meV. Given the increasing interest in quantum information processing using spin and valley states in silicon quantum dots~\cite{Culcer2010}, it is important that the multi-valley level structure of these systems is well characterised experimentally.

\section*{ACKNOWLEDGMENTS}

The authors thank D. Culcer and F. A. Zwanenburg for useful discussions and D. Barber for technical support. This work was supported by the Australian National Fabrication Facility, the Australian Research Council (under contract CE110001027), and by the U.S. National Security Agency and U.S. Army Research Office (under contract W911NF-08-1-0527).

\end{document}